\documentclass{iau}
\usepackage{graphicx}

\title[On producers of cosmic organic compounds] 
{On producers of cosmic organic compounds: exploring the boron 
abundance in lithium-rich K giant stars}

\author[Drake et al.]   %% give here short author list %%
{N. A. Drake$^{1,2}$,  R. de la Reza$^1$, V. V. Smith$^3$, 
K. Cunha$^{1,3}$  }

\affiliation{$^1$Observat\'orio Nacional-MCTIC,  Rua Gen. Jos\'e Cristino 77,
        20921-400, Rio de Janeiro, Brazil,   
    email: {\tt drake@on.br, delareza@on.br, kcunha@on.br} \\
$^2$St. Petersburg State University, Laboratory of Observational Astrophysics, \\
     Universitetski pr. 28, 198504, St. Petersburg, Russia \\
$^3$National Optical Astronomy Observatory, 950 North Cherry Ave., Tucson, AZ 8
5719, USA, \\
     email: {\tt vsmith@noao.edu, cunha@email.noao.edu}}

\pubyear{2017}
\volume{332}  %% insert here IAU Symposium No.
\pagerange{}
\jname{Astrochemistry VII - Through the Cosmos\\
             from Galaxies to Planets}
\editors{M. Cunningham, T. Millar \& Y. Aikawa, eds.}
\begin{document}

\maketitle

\begin{abstract}
The element boron belongs, together with lithium and beryllium, to a known
trio of important elements for the study of evolutionary processes
in low mass stars.  Because B is
the least fragile of this trio to be destroyed in the stellar interiors,
it can be used to test if the  Li enrichment is of planetary origin.
Here, for the first time, boron lines are examined in the
UV for four giants with different degrees of large Li enrichment by
means of observations with the Hubble telescope.
Two main results are found in our study. One is that to first
approximation B abundances appear not to be in excess, invalidating
the planet engulfment mechanism. 
The second one is that the two stars with very large Li abundances present 
emission lines indicating that quite strong active chromospheres are acting
in these very Li-rich giants.
These new results obtained from the UV complement our recent
studies in the mid-IR (de la Reza  et al. 2015) where strong emission-line 
features of organic material
were found in the spectra of some Li-rich stars. 
\end{abstract}

\firstsection % if your document starts with a section,
\section{Introduction}

Following the standard evolution theory, the first ascent red giant stars 
(RGB) must destroy, by means of thermal reactions, almost all of their 
original lithium.  
However, since 1982, when Wallerstein \& Sneden detected the first K giant 
with strong lithium lines, HD\,112127, several Li-rich K giants have been 
detected  having Li abundances, even as large as ten times the 
original Li abundance,  indicating that new sources of $^7$Li are in action.
As an example, Fig.~1 shows the spectrum of the Li-rich giant HD\,19475  
in the region containing the Li\,{\sc i} resonance line at $\lambda$6707.8~\AA.
 
Li, Be, and B are relatively fragile elements which are destroyed by 
$(p,\, \alpha)$ reactions at relatively low temperatures at about 
$2.5\cdot 10^6$, $3.5\cdot 10^6$, and $5.0\cdot 10^6$~K, respectively.
Since 1982 the nature of these Li-rich red giant stars have been 
involved in a problem known as the ``lithium puzzle''. 

\begin{figure}[t]  %fig 1
% \vspace*{2.0 mm}
\begin{center}
 \includegraphics[width=2.84in]{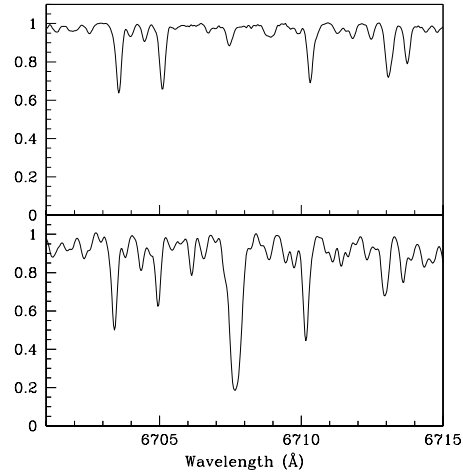}
 \caption{Spectrum of the Li-rich giant HD\,787 (bottom panel) in 
         comparison with a normal K giant star $\epsilon$~Vir (upper panel) 
         in the region of the Li\,{\sc i} line at 6707.8~\AA.}
\label{fig1}
\end{center}
\end{figure}

Recently we have discovered in the near-IR (de la Reza et al. 2015)  the 
presence of complex organic compounds in the shells of Li-rich K giant stars
(Fig.~2). 
These stars present large mass loss transforming them into sources of 
organic material in the Galaxy. 

\begin{figure}[h]  %fig2
\begin{center}
 \includegraphics[width=3.14in]{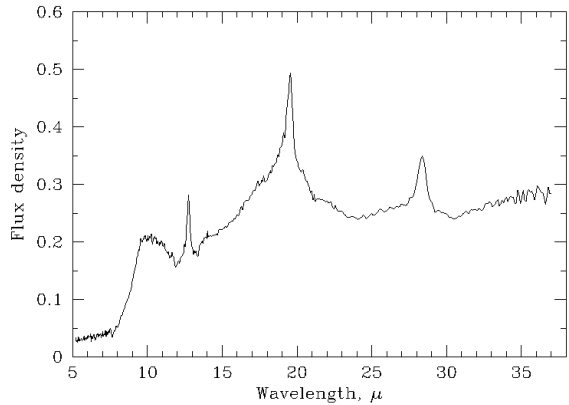}
 \caption{Spitzer mid-IR spectrum of the Li-rich K giant star PDS\,68  
         between 5 and 38~$\mu$m presenting the UIE features superposed 
         on a strong continuum emission  (de la Reza et al. 2015). The spectrum was taken from the Cornell Atlas of {\
it Spitzer}/IRS sources (CASSIS, Lebouteiller et al. 2011).  }
\label{fig2}
\end{center}
\end{figure}

Beryllium abundances in lithium-rich giants, including the star HD\,787,
were studied by Castilho et al. (1999) and Melo et al. (2005). 
The authors have  found that 
Be abundances in these stars are low, suggesting that Be is very depleted 
($>$90\%) from the initial value of the Populaton~I stars.     

The nature of these stars is the object of recent studies. Here, it is explored 
for the first time the boron lines in the spectra of these objects in the UV
spectral region with the aim to infer their main properties.

\section{Observations and stellar parameters}

We observed four Li-rich K giants HD\,787, HD\,9746, HD\,3953, and 
HD\,112127 with the Hubble Space Telescope and Space Telescope Imaging 
Spectrograph STIS with grating G230M having FWHM\,=\,30~km\,s$^{-1}$
and spectral region centered at the resonant B\,{\sc i} lines at 
$\lambda$2496.8 and $\lambda$2497.7~\AA. % (Proposal ID\,8612).
These four spectra are shown in Fig.~3.  

\begin{figure*}[h]  %fig3
\begin{center}
 \includegraphics[width=3.94in]{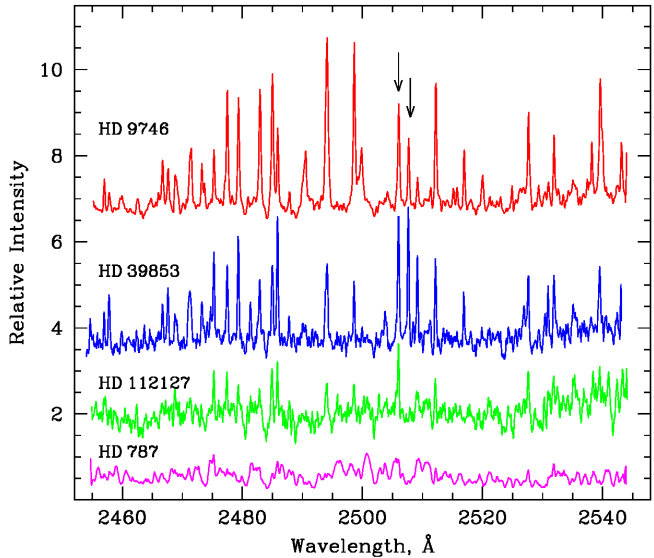} 
 \caption{The HST/STIS UV spectra of four giant stars in the  
 region of the B\,{\sc i} lines at $\lambda 2496.8$ and $\lambda 2497.7$~\AA.
 From top to bottom: HD\,9746 ($\log\varepsilon{\rm (Li)}=3.75$), HD\,39853 
 ($\log\varepsilon{\rm (Li)}=2.85$), HD\,112127 
 ($\log\varepsilon{\rm (Li)}=2.7$),
 and HD\,787 ($\log\varepsilon{\rm (Li)}=1.8$).
The UV spectra of the Li-rich giants HD\,9746 and HD\,39853 are very 
complex and rich in strong emission lines, mainly of Fe\,{\sc ii}.
As can be seen, the intensity of the emission lines correlates with 
the Li abundance. 
Very strong emission lines at $\lambda 2507$ and $\lambda 2509$~\AA\ 
marked by arrows are observed in the spectra of HD\,9746 and HD\,39853. 
These two lines are the Fe\,{\sc ii} fluorescence lines pumped by 
H\,Ly$\alpha$ (Johanson \& Hamann 1993) pointing to the existence of very 
strong Ly$\alpha$ emission. 
Star HD\,787 shows a different spectrum probably due to its binary nature.
The spectra are arbitrary shifted in the relative intensity axis.  }
\label{fig3}
\end{center}
\end{figure*}

The atmospheric parameters of the stars, effective temperature 
($T_{\rm eff}$), surface gravity ($\log g$), and lithium abundance 
($\log\varepsilon{\rm (Li)})$  
(where $\log\varepsilon{\rm (X)} =\log (N_{\rm X}/N_{\rm H}) + 12.0$),
are shown in Table~1.

\begin{table}[t]
 \begin{center}
 \caption{Atmospheric parameters of observed Li-rich K giant stars.}
 \label{tab1}
\begin{tabular}{lcccc}
\hline
Star &$T_{\rm eff}$, K & $\log g$ &$\log\varepsilon({\rm Li})$ & Reference\\
\hline
HD\,787$^*$&  4220        & 1.50    &  1.8  & (1) \\
HD\,9746   &  4400        & 2.30    & 3.75  & (2) \\
HD\,39853  &  3900        & 1.16    & 2.85  & (3) \\
HD\,112127 &  4340        & 2.10    & 2.7   & (4) \\
\hline
\end{tabular}
\end{center}
\vspace{1mm}
\begin{center}
{\it References:}
(1) Brown et al. (1989);\\ (2) Balachandran et al. (2000);\\
(3) Gratton \& D'Antona (1989);\\  (4) Wallerstein \& Sneden (1982)\\
{\it Notes:}
$^*$The star HD\,787 is a proper-motion binary\\ (Mason et al. 1999, 
Eggleton \& Tokovinin 2008) %Frankowski et al. 2007)
\end{center}
\end{table}

\section{Results and conclusions: how to explain the Li puzzle?}
\vspace{1mm}
\begin{itemize}
\item  The boron abundances of the Li-rich giants appear to be  low and 
in none of the cases presents an excess. This result is not compatible 
with any scenario of engulfment of planets in order to produce Li-rich K giant 
stars. 
\item  In de la Reza et al. (2015) we propose a qualitative scenario to try 
to explain the ``Li puzzle'' in K giant stars. Here, a mass mixing produced 
by the diminishing of the stellar core rotation in the RGB phase, 
reaches the surface layers of the stars forming eventually the observed 
circumstellar shell. At the same time, this mixing can, in some short 
time episodes, move the $^7$Be produced in the internal H-burning region 
rapidly to the external convective zone where it is transformed 
into new $^7$Li. 

\item  Privitera et al. (2016) suggested that a strong magnetic field in 
a red giant star may be a sign of a past planet engulfment.  
It is interesting to note that our observed very Li-rich K giant HD\,9746 
has an important magnetic field detected by Auri\`ere et al. (2015). 
We found in this work that this K giant star can be a descendant of an Ap 
star on the main sequence known to have very high Li abundance. 
Because HD\,9746 has a relatively high rotation of $v\sin i =8.7$~km\,s$^{-1}$,
  
its magnetic field can have two origins: one fossil from the Ap ancestor 
and one produced by a rotationally induced dynamo. 
Goriely (2007) suggested that high lithium abundance in magnetic Ap stars 
may be produced by the spallation reactions in the regions of the magnetic 
poles, where accelerated protons and $\alpha$-particles destroy CNO nuclei 
and produce lithium.

\item  Even if the direct relation of the Li enrichment phenomenon and 
existence of a magnetic field is not yet well understood, there exists a 
strong correlation between stellar activity and high Li abundances in 
red giant stars (Fekel \& Balachandran 1993, Drake et al. 2002).
\end{itemize} 

\begin{figure}[h]  %fig4
\begin{center}
 \includegraphics[width=2.64in]{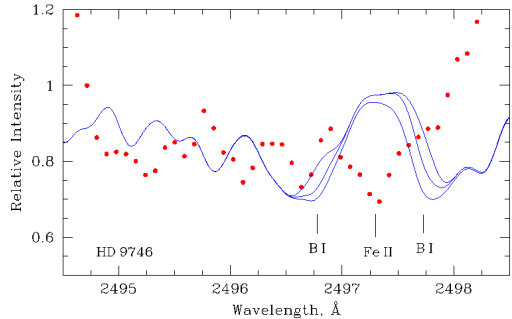} 
 \caption{Fit of the observed spectrum of HD\,9746 in the spectral region 
  around the resonance doublet of B\,{\sc i} at $\lambda 2496.8$ and 
  $\lambda 2497.7$~\AA\ in the spectrum of HD\,9746. 
 {\it Points}: observations.  {\it Lines}: synthetic spectra computed 
 using the recent version of {\sc moog} code (Sneden 1973) for the boron 
 abundances of $\log\varepsilon{\rm (B)}=0.60$, 2.60, and 4.60 
 corresponding to the [B/Fe] = --2.0, 0.0, and +2.0. 
 Strong contribution from the chromosphere causing veiling in this UV region, 
 makes it difficult to determine the B abundance with high precision.  
 Nevertheless, a direct comparison with theoretical B\,{\sc i} lines 
 shows that no expected B excess is present. 
  The line lists were taken from the VALD Database, Duncan et al. (1998), 
 and Cunha et al. (2000). }
\label{fig4}
\end{center}
\end{figure}

{\it Acknowledgements.} N.A.D. acknowledges IAU travel grant, FAPERJ, 
Rio de Janeiro, Brazil, for Visiting Researcher grant E-26/200.128/2015, 
and the Saint Petersburg State University for research grant 6.38.335.2015.
%\vspace{-1mm}

\end{document}